\documentclass[twocolumn]{article}

\usepackage[english]{babel}
\usepackage[letterpaper,top=2cm,bottom=2cm,left=3cm,right=3cm,marginparwidth=1.75cm]{geometry}
\usepackage{abstract}
\usepackage{amsmath}
\usepackage{graphicx}
\usepackage[colorlinks=true, allcolors=blue]{hyperref}
\usepackage{authblk}

\usepackage{caption}
\title{Charge sensitivity in the transmon regime}
\date{}  % This removes the date
\author[1,2,\textsuperscript{\dag}]{Rocio Gonzalez-Meza}
\author[1,\textsuperscript{\dag}, *]{Vito Iaia}
\author[3]{Anika Zaman}
\author[3]{Hiu-Yung Wong}
\author[1]{Yujin Cho}
\author[1]{Kristin Beck}
\author[1]{Yaniv J. Rosen}

\affil[1]{Lawrence Livermore National Laboratory, Livermore, CA 94550}
\affil[2]{William H. Miller III Department of Physics, Johns Hopkins University, Baltimore, MD 21218}
\affil[3]{Electrical Engineering, San Jose State University, San Jose, CA 95192}
\affil[$\dagger$]{These authors contributed equally: R. Gonzalez-Meza, V. Iaia.}
\begin{document}
\twocolumn[
\begin{@twocolumnfalse}
\maketitle
\begin{center}
*\texttt{iaia1@llnl.gov}
\end{center}
\begin{abstract}

   Transmons are widely adopted in quantum computing architectures for their engineered insensitivity to charge noise and correspondingly long relaxation times. Despite this advantage, transmons often exhibit large fluctuations in dephasing times across different devices and also within qubits on the same device. Existing transmon qubits are assumed to be insensitive to charge noise. However, very little recent attention has been paid to the dependence of dephasing on the local charge environment. In this study, we see fluctuations in the dephasing time, $T_{\phi}$, which correlate to charge offset. While charge offset fluctuations are slow, parity switches are fast processes tied to the charge offset and can affect $T_{\phi}$ in Ramsey experiments. We implement a protocol to detect parity switching events using single-shot methods, which are interleaved within a Ramsey measurement. We find that events that remain in the same parity state have a higher $T_2$ than measurements averaged over both parities. Our results show that transmons can be limited by charge-noise, even with $E_\textup{J}/E_\textup{C} \approx 50$. Consequently, parity flip rates must be considered as a device characterization metric.

\end{abstract}
\end{@twocolumnfalse}
]

\section{Introduction}
%% JJ-based devices are susceptible to charge noise 
Transmon qubits are Josephson junction-based devices whose non-linearity provides anharmonic energy levels that can be controlled \cite{krantz2019quantum}. The Cooper Pair Box \cite{Bouchiat1998}, a precursor of the transmon, was plagued by charge noise. Charge noise arises from fluctuations in the local electrostatic environment of the device \cite{Kenyon2000, Dutta1981, Mueller2019}. These fluctuations couple to the qubit and induce variation in energy level spacings, also known as charge dispersion. This leads to dephasing noise, and ultimately limits the coherence times of the quantum states used in computations. Koch \textit{et al.} introduced the transmon qubit in 2007 to mitigate the charge sensitivity inherent to the Cooper pair box design~\cite{koch2007}. The transmon achieves this by increasing the ratio of the Josephson energy $E_\textup{J}$ to the charging energy $E_\textup{C}$, which exponentially suppresses charge dispersion by $\exp{(-\sqrt{8E_\textup{J} / E_\textup{C}})}$. However, this comes at the cost of reduced anharmonicity. In Ref.~\cite{koch2007}, a ratio of $E_\textup{J} / E_\textup{C} \approx 50$ was proposed as an optimal compromise between charge noise suppression and qubit addressability. The device we study in this article operates in this regime, where charge dispersion is expected to be strongly suppressed but not entirely eliminated. Consistent with this expectation, we measure a residual charge dispersion of $\sim 6$~kHz.

There are two timescales for charge offset fluctuations over a Josephson junction.  Charge sensitive qubits typically have a slow drift of $\approx 0.5 \ e / \textup{hr} $ \cite{christensen2019, tennant2022, wilen2021}. In addition, faster charge offset fluctuations can occur due to quasi-particle poisoning (QP). QPs originate from broken Cooper pairs. When a quasiparticle tunnels across a Josephson junction, the charge offset changes by $e$, and the total number of electrons on each conductor changes by one.  If the number of electrons on a conductive island is even, the superconductor is in an \textit{even} parity state. A quasiparticle tunneling event will change an \textit{even} parity to \textit{odd} and vice versa. This fast parity switching process can be seen on many devices.\cite{Riste2013,serniak2018,wilen2021,diamond2022parity,iaia2022}

Recent studies have highlighted the impacts of these noise mechanisms on qubit coherence and their implications for device optimization \cite{Papic2024}. The tunneling of QPs results in two distinct parity states of the qubit. The device can randomly alternate between these two states, changing the frequency of the qubit transition, resulting in telegraph noise \cite{Riste2013, tennant2022, serniak2018, christensen2019}. In this study, we experimentally demonstrate how QP-induced parity switches impact the dephasing time of a single transmon qubit fabricated from tantalum.

This article is structured as follows. In Section~\ref{sec:methods}, we describe the methodology used to detect and analyze parity flips, including the spectroscopy experiments used to identify the charge-parity bands. Section~\ref{sec:results} presents the results of our study, highlighting the effects of parity switches on qubit coherence. We also explore the impact of charge-parity noise on dephasing times to evaluate their sensitivity to QP tunneling events. In Section \ref{sec:discussion}, we discuss the broader implications of our findings, as well as the limitations of our approach. Finally, in Section \ref{sec:conclusion}, we summarize the key results.

\section{Methods}\label{sec:methods}
\begin{figure}[hbt!]
    \centering
    \includegraphics[width=\linewidth]{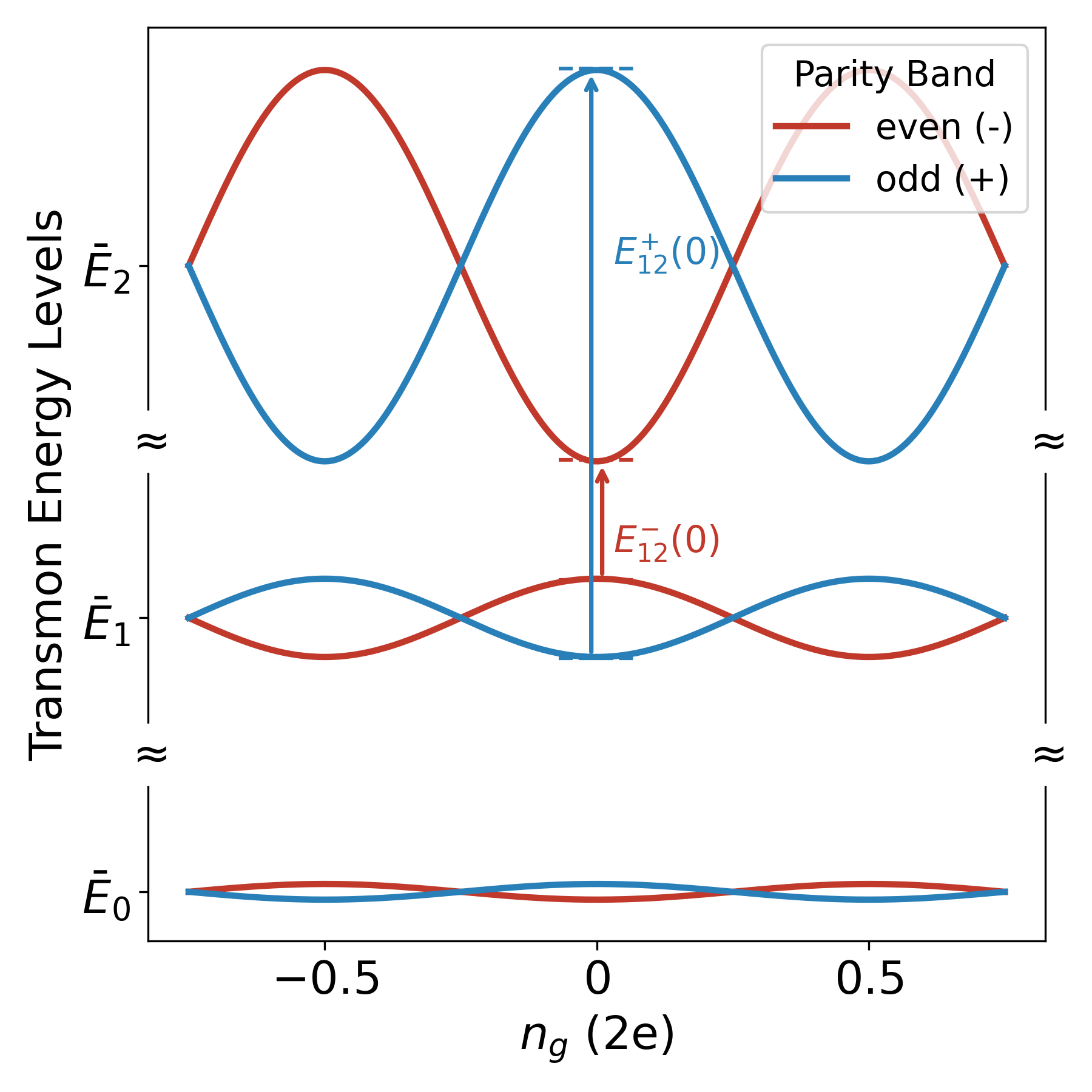}
    \caption{A diagram of a transmon's three lowest energy levels as a function of charge offset $n_g$. The colored traces (not to scale) indicate even and odd parity states. As $n_g$ changes, the separation between parity states changes sinusoidally. This difference becomes more prominent in higher energy levels.}
    \label{fig:parity-bands}
\end{figure}

The transition frequency between levels $i$ and $j$ of a transmon qubit is approximately described by % 

\begin{equation}
    f_{ij}^{\pm} \approx \bar{f}_{ij} \pm \epsilon_{ij} \cos{(2\pi n_g)},
    \label{eq:frequency-transitions}
\end{equation}
where $\bar{f}_{ij}$ is the average transition frequency, $n_g$ is the dimensionless gate charge (in units of $2e$), and $\epsilon_{ij}$ is the amplitude (or half the frequency difference between the two parities) of the charge dispersion (Figure \ref{fig:parity-bands}). The $\pm$ represents the parity-dependent shift in frequency. However, since we can only measure $n_g$ modulo $e/2$, we can detect parity flips but cannot determine the exact parity. We associate $f_{ij}^{+}$ with a higher frequency and $f_{ij}^{-}$ with a lower frequency, each representing a different parity state.

\begin{figure*}[hbt!]
    \centering
    \includegraphics[width=\textwidth]{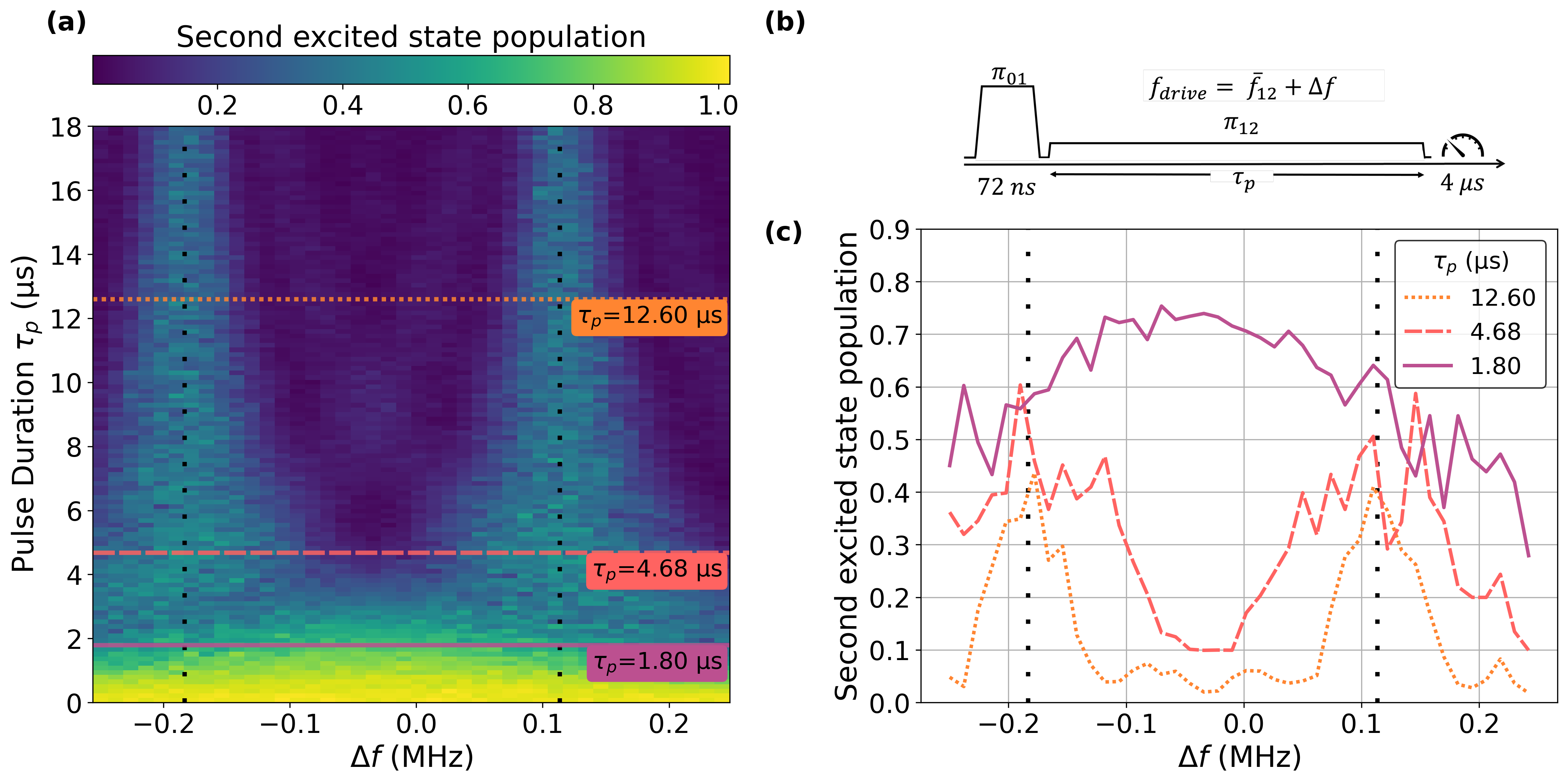}
    \caption{Parity frequencies detection and pulse duration optimization using high-resolution spectroscopy. (a) The population of the second excited state, $P_{|2\rangle}$, is measured as a function of the drive frequency offset $\Delta f$ and pulse duration $\tau_p$. Longer pulses reveal the presence of two distinct parity states at resonant frequencies $f_{12}^{\pm}$. The frequencies at the two parities, $f_{12}^{\pm}$, are marked by the dotted vertical lines.
    (b) Single-shot sequence for high-resolution spectroscopy: We vary the frequency and duration of the applied $\pi_{12}$ pulse as $f_{\text{drive}} = \bar{f}_{\text{12}} + \Delta f$ and $\tau_p$, respectively, where $\bar{f}_{\text{12}}$ is the nominal frequency of the 1-2 transition. The pulse amplitude is adjusted for each duration $\tau_p$ to ensure a $\pi$-rotation. Parity states are equally probable, meaning we expect the peaks to reach 0.5. For very long pulses the maximum population at the resonant frequencies decreases due to decoherence effects. (c) Spectra at different pulse durations. As the pulse duration increases, the two peaks become more distinct.}
    \label{fig:frequency-detection-and-pulse-optimization}
\end{figure*}

We define the frequency offset between the two parity bands as:

\begin{equation}\label{eq:charge_dispersion}
    \Delta_{ij} (n_g) \equiv \frac{|f_{ij}^{+} - f_{ij}^{-}|}{2} = \epsilon_{ij} \cos{(2\pi n_g)}.
\end{equation}

 Parity switching rates have been observed in the literature on the order of 0.1 - 100 kHz \cite{sun2012measurements,serniak2018,Riste2013}, however, improvements in qubit designs can push this number down to $\sim$1~mHz \cite{pan2022engineering,iaia2022,kamenov2023suppression,kurter2022quasiparticle}. 

Fluctuations of the charge offset occur on much slower timescales, typically on the order of minutes to hours \cite{tennant2022, wilen2021}, allowing us to treat $n_g$, and therefore $\Delta_{ij}$, as quasi-static during experiments that probe parity dynamics.

\subsection{Resolving the parity states}
% Rewrite this from the experimenrtal perspective: what did we measure? 
For our device $E_J/E_C  \approx  50$, and the charge dispersion for the qubit's 0-1 transition is expected to be $\sim 6 \  \textup{kHz}$, which is below the resolution of our measurement protocol. Instead we leverage the larger charge dispersion of the 1-2 transition, which we measured to be $\sim150~\textup{kHz}$ \cite{tennant2022}. Our typical $\pi$-pulse durations are around 72 ns, which do not provide enough frequency resolution to distinguish the parity states. To address this limitation, we performed spectroscopy of the (1-2)~transition for different $\pi_{12}$~pulse durations, as shown in Figure \ref{fig:frequency-detection-and-pulse-optimization}. This high-resolution spectroscopy serves a dual purpose: (1) it identifies the resonant frequencies of the two parities, and (2) it determines the optimal pulse duration for parity detection using single-shot measurements. 

% What happens Increase pulse length ?
Normally we use a 72 ns $\pi$-pulse to excite the 1-2 transition. This pulse duration transfers $100\%$ of the population of the first excited state to the second excited state, revealing a broad peak (See Fig. \ref{fig:frequency-detection-and-pulse-optimization} (a)). As we increase the pulse length the frequency dependence of the population reveals two peaks. The sharpness of these peaks increases with the pulse duration. Longer pulses enable better frequency resolution.  In our case we have peaks with a separation of 300 kHz, which are resolved with pulses longer than $\sim 3 \mu$s. The two peaks have a reduced population, close to 50\%, because during a measurement, there is an equal probability for each parity state to be populated. At very long durations the effects of relaxation and dephasing become relevant ($T_{2}^{*} = 23~\mu$s), affecting the population. This can be observed in Fig. \ref{fig:frequency-detection-and-pulse-optimization} (c), where we see that longer pulses provide peak distinction, but the population is reduced. For example, $\tau_p = 4.7 \mu $s has peaks close to 0.5 but $\tau_p = 12.6 \mu$s looks slightly reduced. We chose an optimal pulse length that allows for maximal peak separation but with an excitation probability closest to 0.5. 

Since the peak separation changes over time as a function of offset charge, the optimal pulse length changes as well. We perform this protocol prior to any measurement that requires parity distinguishability.

\subsection{Parity Flip Detection}
\begin{figure}[ht!]
    \centering
    \includegraphics[width=\linewidth]{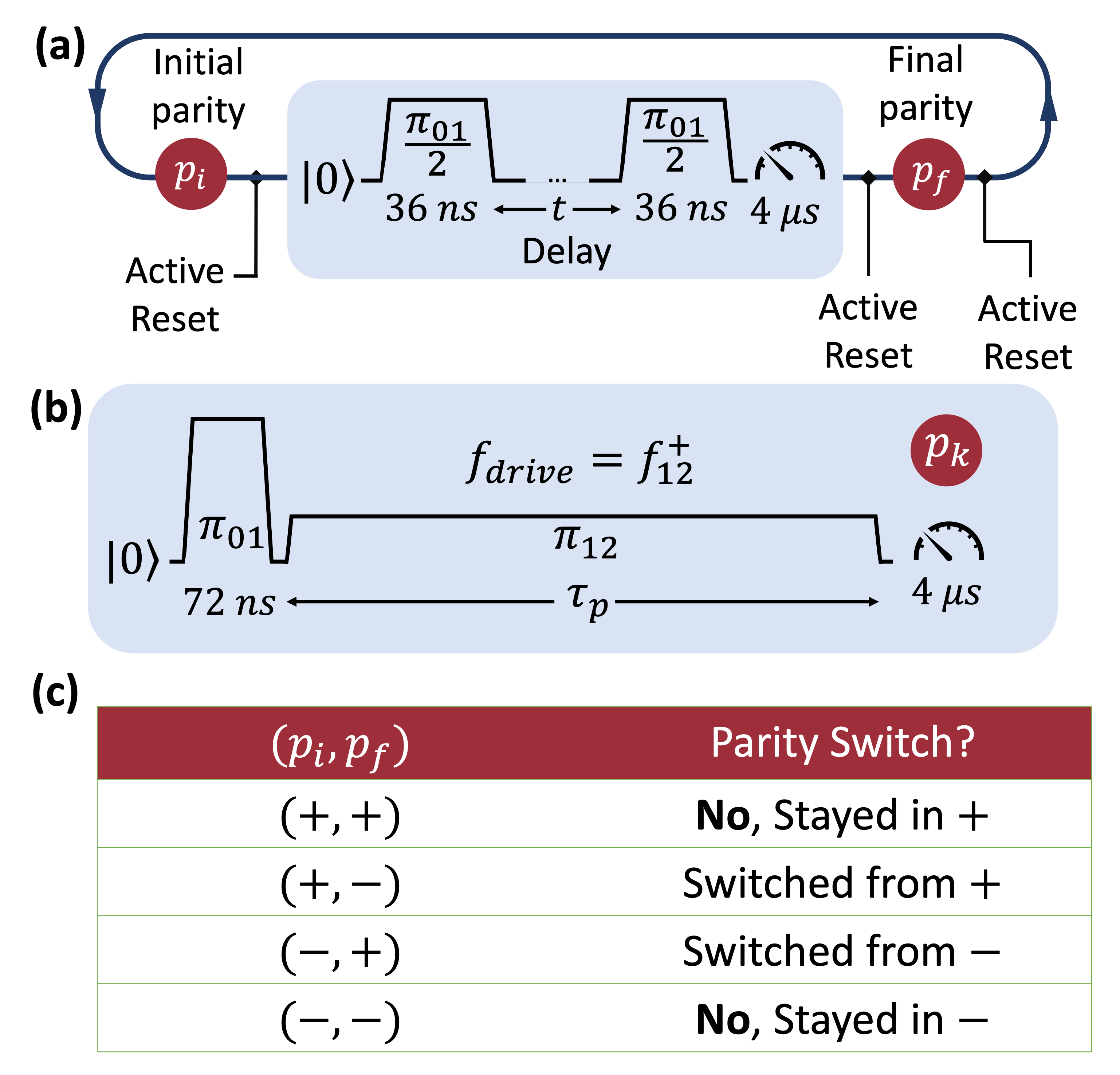}
    \caption{
        \textbf{Embedded parity detection protocol.}
        (a) Schematic of the Ramsey decay experiment with embedded parity detection: For each shot at a fixed delay time, parity detection sequences are performed at the beginning ($p_i$) and end ($p_f$) of the Ramsey sequence. This allows identification of parity flips during the experiment by comparing $p_i$ and $p_f$.
        (b) Detailed view of the parity detection sequence, labeled as $p_k$ ($k = i, f$): A selective pulse at $f_{12}^{+}$, and optimal duration $\tau_p$ is applied, and the outcome determines the parity. Detection of the $|2\rangle$ state indicates "$p_k = + $" parity; otherwise, the system is in the opposite parity "$p_k = - $". The flow diagram illustrates the decision process based on the measurement outcome.
        (c) Table of possible $(p_i, p_f)$ combinations: Each pair indicates whether a parity flip occurred during the Ramsey sequence.
    }   
    \label{fig:parity-protocol-diagram}
\end{figure}

By tuning the drive frequency to either $f_{12}^{+}$ or $f_{12}^{-}$, we can selectively excite transitions with a specific parity. A parity flip is identified when consecutive measurements targeting the same transition yield different outcomes. For instance, if the first measurement at $f_{12}^{+}$ fails to excite the transition, the system is in the opposite parity state, which corresponds to $f_{12}^{-}$. If a parity flip occurs, a subsequent measurement will successfully excite the $f_{12}^{+}$ transition. This method allows us to identify parity changes during the experiment.
This method depends on parity flips occurring at slow rates compared to the experiment.
Parity detection is performed in the 1-2 manifold, while characterization experiments, such as Ramsey decay, are conducted in the 0-1 manifold. Figure \ref{fig:parity-protocol-diagram} illustrates a typical experimental sequence, where we interleave parity detection between every single shot measurement. A parity detection typically takes on the order of 10 $\mu$s, due to the pulse duration and readout time. After each experimental step, an active reset is applied to return the qubit to its ground state, which takes approximately 10 $\mu$s due to readout. The Ramsey experiment time is dominated by the delay and it can take as long as $100 \mu$s. We measured parity flip rates on the order of 1 kHz, which is unlikely to have more than one parity flip during this sequence. Each delay time was measured a thousand times. The total span per delay point was on the order of seconds, and therefore encompassed many parity flips. A full Ramsey measurement took approximately five minutes, which can be compared to the rate of charge offset jumps, which occurred approximately once per hour.  
% too many approx! change to ~

By using embedded parity detection, we categorize measurements into parity-flipped and unflipped cases. Unflipped cases are further grouped by their initial parity state, which we represent by the frequencies $f_{12}^{\pm}$ used to excite them.  

\section{Results}\label{sec:results}

\subsection{Effect of parity flips on qubit coherence}
\begin{figure}[h!]
    \centering
    \includegraphics[width=\linewidth]{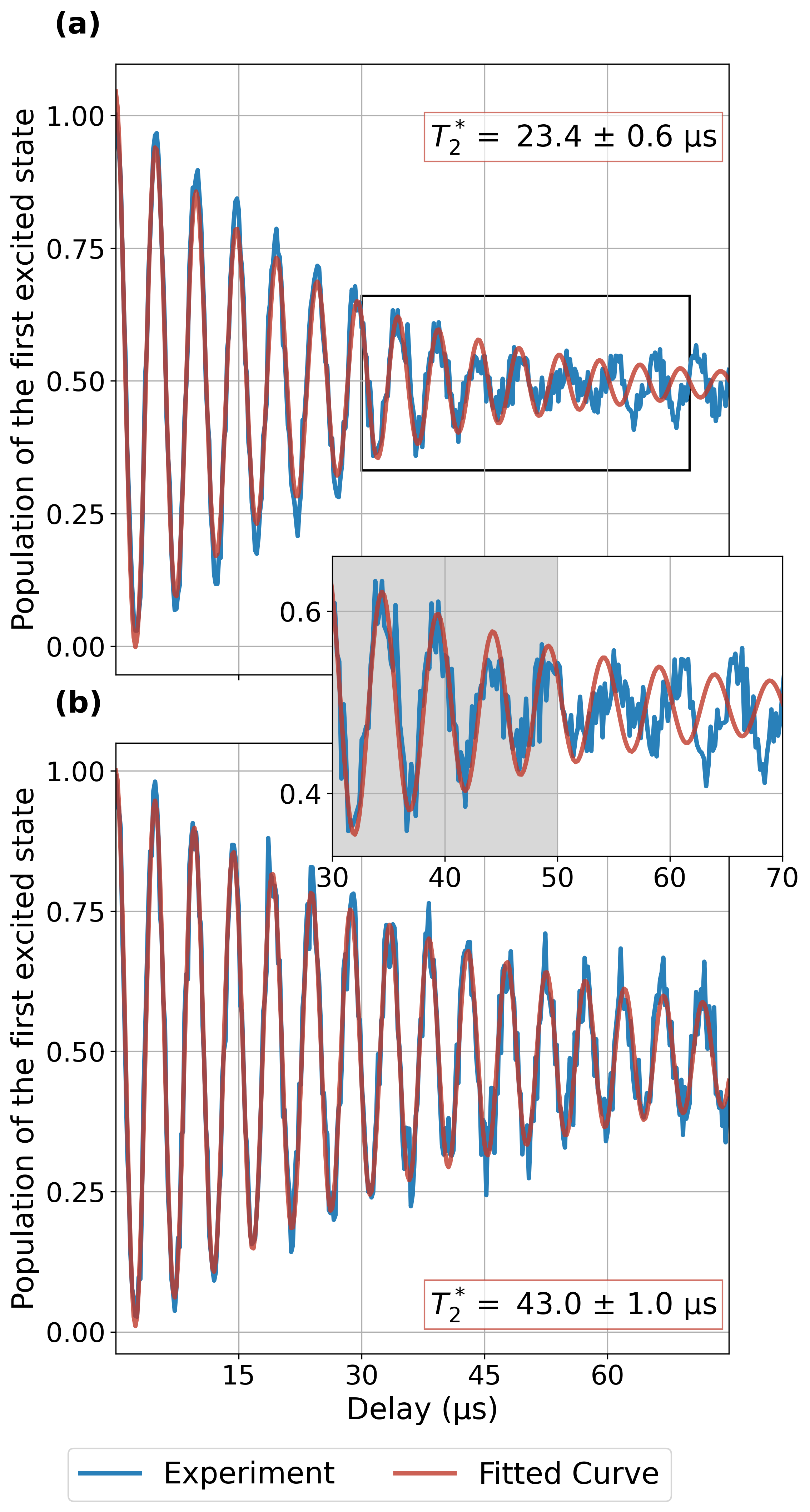}
    \caption{Ramsey decay experiment on the 0-1 subspace with embedded parity detection measurements. (a) Experiments agnostic to parity flip events exhibit $T^{*}_{2}$ = 23.4 +/- 0.6 $\mu$s. The inset seems to exhibit a beating pattern indicative of a signal containing two characteristic frequencies; the gray-shaded region highlights the time interval where the experimental data and a fitted $T_2^*$ exponential decay curve are in good agreement. Outside this region, the signal and fit begin to diverge. (b) We post-select Ramsey decay experiments with no parity flips, which begin at parity $f^{+}_{12}$, and find that $T^{*}_{2}$= 43.0 +/- 1.0 $\mu$s.}

\label{fig:parity-flip-single_experiment}
\end{figure}

Figure \ref{fig:parity-flip-single_experiment} shows a Ramsey decay experiment with embedded parity detection measurements. For the unsorted dataset (Figure \ref{fig:parity-flip-single_experiment}(a)), the Ramsey oscillations result in a dephasing time of $T_2^* = 23.4 \pm 0.6~\mu$s. In contrast, the measurements that stayed in the $f_{12}^{+}$ parity (Figure \ref{fig:parity-flip-single_experiment}(b)) exhibit a substantially longer dephasing time $T_2^* = 43.0 \pm 1.0~\mu$s. Furthermore, the signal in Figure \ref{fig:parity-flip-single_experiment} (a) seems to display a beating pattern (inset), which arises due to the presence of two characteristic frequencies associated with parity flips.

\subsection[Impact of Charge-Parity Noise on T2*]{Impact of Charge-Parity Noise on $T_2^*$}
\begin{figure*}[hbt!]
    \centering
    \includegraphics[width=\linewidth]{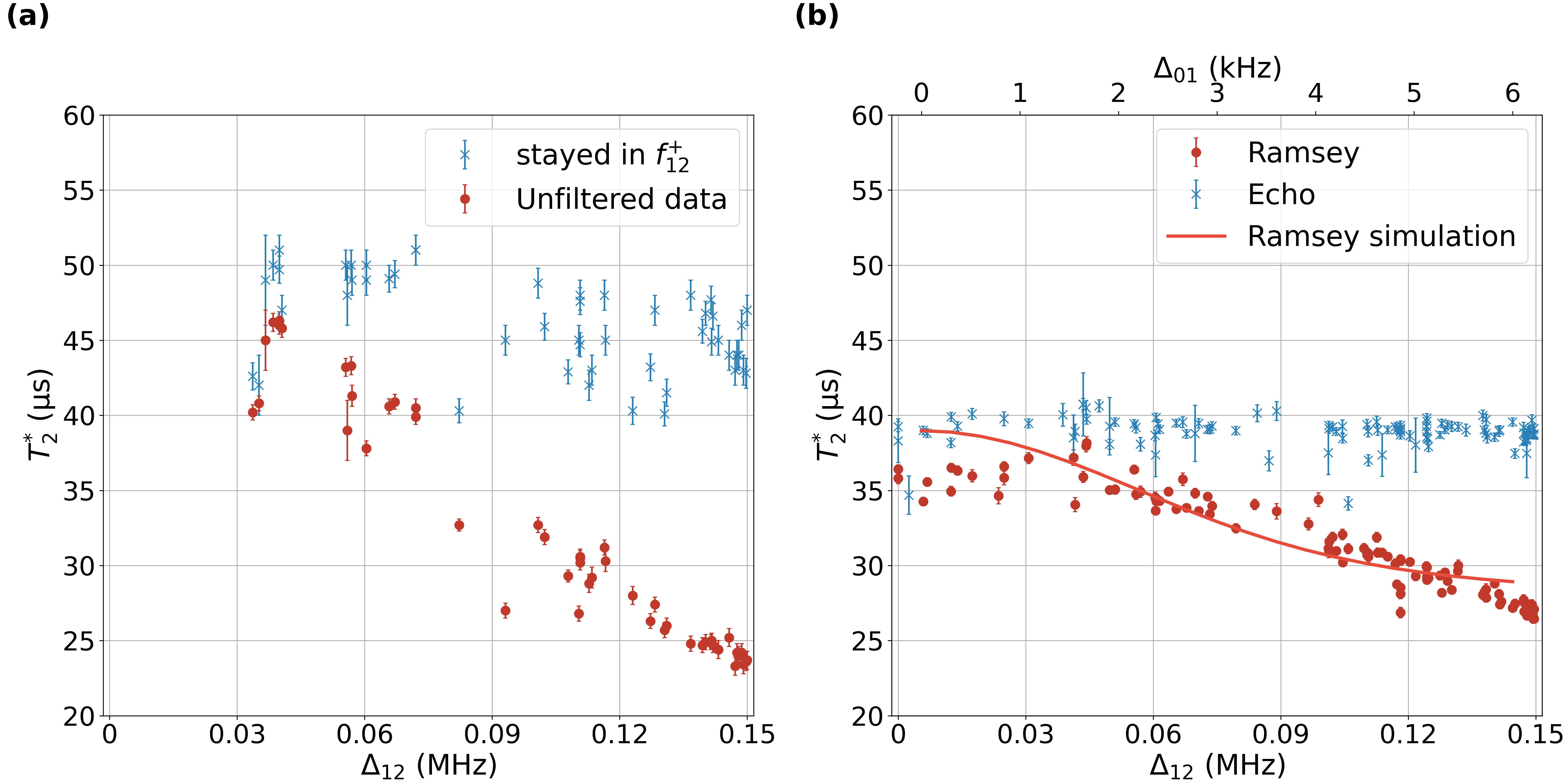}
    \caption{Dephasing time $T^{*}_2$ extracted from Ramsey experiments in the 0-1 subspace. (a) Results from experiments performed with embedded parity measurements. We observe no variation in $T^{*}_2$ with charge offset for events with no parity switch (blue cross). However, Measurements with parity flip exhibit a decrease in $T^{*}_2$ with increasing charge offset (red dots). (b) Results from Ramsey measurements with (blue cross) and spin echo (red dots). No parity detection steps were included. The red line represents the simulation of $T^{*}_2$ times affected by charge-parity noise for different charge-offsets in the $0-1$ manifold ($\Delta_{01}$, top x-axis).}
    \label{fig:simulation-vs-experiment}
\end{figure*}

To investigate the effects of charge-parity noise on coherence, we performed repeated Ramsey decay experiments with embedded parity detection steps. The measurement events were sorted into two separate categories: with and without parity flips, with the probe frequency at $f_{12}^{+}$. The dephasing times, $T_2^*$, were then extracted for both cases. The results are summarized in Figure \ref{fig:simulation-vs-experiment}(a). For unsorted data, $T_2^*$ decreases as the charge offset increases. In contrast, measurements without parity flips consistently exhibit longer dephasing times. These results confirm that charge-parity noise is a significant source of decoherence in the system.

To further explore the impact of charge-parity noise, we compared dephasing times from Ramsey and spin-echo experiments, both performed using the phase method \cite{Wong2024} for detuning-insensitive measurement. This approach leverages virtual Z-rotations to implement the evolution rather than relying on physical idling, making the results robust against fluctuations in the offset charge $n_g$. The spin-echo sequence, which incorporates a refocusing $\pi$-pulse midway through the evolution, effectively cancels out dephasing from slow and quasi-static noise sources, thereby isolating the contribution from faster noise processes.

The estimated dephasing times from both Ramsey (blue crosses) and spin-echo (red dots) experiments are illustrated in Figure~\ref{fig:simulation-vs-experiment}(b). Typically, qubit measurement dephasing times vary randomly over the course of a day. However, here we observe a clear trend: Ramsey dephasing times decrease monotonically with increasing charge dispersion, reflecting the influence of charge-parity noise. In contrast, the dephasing times from spin-echo experiments remain constant across varying charge offsets. This behavior demonstrates that the spin-echo technique mitigates the effects of charge-parity noise.

To substantiate our understanding of the effect of charge-parity noise on qubit coherence, we simulated the Ramsey decay of our device subjected to parity flips (See appendix \ref{sec:App_Lindblad}). To do this, we simulated the qubit evolution with a parity flip at some time $t$. Because the total evolution time is much less than the parity flip time, we averaged over the effect of a parity flip occurring at every time step. We obtained $T_2^*$ from the exponentially decaying envelope of the simulated sequence. In order to simulate varying charge dispersion, we used parity flip frequencies $\Delta_{01}$ ranging from 0 to 6 kHz. The simulated $T_2^*$ as a function of charge offset is plotted (red trace) in Figure \ref{fig:simulation-vs-experiment}(b).

\section{Discussion}\label{sec:discussion}

From a signal analysis perspective, parity flips introduce two characteristic frequencies into the Ramsey signal, resulting in a beating pattern when averaged over many shots \cite{Martinez2023}. In practice, some of these measurements remained in $f_{01}^+$, others in $f_{01}^-$, and some experienced flips between the two. The superposition of these slightly different frequencies leads to an apparent reduction in the exponential decay envelope when the data are fit to a single-frequency model (Fig. \ref{fig:parity-flip-single_experiment}). This effect is more pronounced at higher charge dispersions, but even when the beating pattern is not directly visible due to the duration of the experiment, the net result is a reduced $T_2^*$. By filtering and analyzing only those experiments that remained in a single parity, we recovered a single-frequency signal which would represent the ideal value ($T_2^{*\text{ideal}}$) obtained in the absence of parity flips.

Our data in Figure~\ref{fig:simulation-vs-experiment}(a) and (b) demonstrate that the dephasing time $T_2^*$ decreases monotonically with increasing charge dispersion. In contrast, $T_2^*$ extracted from experiments that remain in a single parity state remain stable across varying charge dispersions. This strongly suggests that fluctuations in $T_2^*$ are primarily driven by charge-parity noise.

Ramsey decay and spin-echo measurements are sensitive to different spectral components of the noise affecting the qubit \cite{Hahn1950,Ban1998,Viola1998}. Ramsey experiments are susceptible to low-frequency noise, such as quasistatic fluctuations and $1/f^2$ components, while spin-echo sequences filter out these slow fluctuations and are sensitive to higher-frequency noise processes. Our results show that $T_2^{\text{Echo}}$ measurements are independent of charge dispersion (Fig. \ref{fig:simulation-vs-experiment} (b)), which supports the hypothesis that the low-frequency noise components affecting $T_2^*$ originate from charge-parity fluctuations. The latter is true provided that the timescale between parity flips ($\sim$1 ms) is much longer than the duration of a single shot in the spin-echo sequence ($\sim$100~$\mu$s). However, if parity flips become as frequent as, or faster than, a single shot in the experimental sequence, the noise spectrum shifts to higher frequencies, and both Ramsey and spin-echo measurements become similarly sensitive to the noise. Another consequence of faster parity switches is that measuring parity flips becomes unreliable, as multiple flips can occur within a single run.

While charge-parity noise explains most of the changes we see in $T_2^*$ with charge dispersion, our data also show other trends that this mechanism cannot account for. For instance, Figures~\ref{fig:simulation-vs-experiment}(a) and (b) reveal a difference of about $5\,\mu$s in $T_2^*$, even though both measurements were taken on the same device a few months apart. This suggests that, although our protocol captures short-term changes in $T_2^*$ (over hours or days), there are additional, slower processes affecting the qubit's coherence over longer timescales.

It should also be noted that in systems where $T_2^*$ is intrinsically short due to other decoherence mechanisms (e.g., energy relaxation), the impact of charge-parity noise is less pronounced. In such cases, the improvement in $T_2^*$ achieved by filtering out parity flips is often within the uncertainty of the measurement itself. As detailed in Appendix~\ref{subsec:charge_parity_noise} and Fig.~\ref{fig:T2star_vs_T2ideal}, the significance of charge-parity noise diminishes as $T_2^*$ decreases.

Overall, these observations provide a comprehensive understanding of when and how charge-parity noise is a relevant and measurable phenomenon in transmon devices operating near the transmon regime. Our findings highlight the importance of considering charge-parity noise in the design and characterization of superconducting qubits, particularly as devices approach longer coherence times and as experimental protocols become more sensitive to subtle noise mechanisms.

\section{Conclusion}\label{sec:conclusion}

% This paragraph summarizes overall what we did: 
In this work, we have developed an experimental protocol to directly probe the effects of charge-parity noise, induced by quasiparticle tunneling, on transmon qubit coherence times. By resolving parity frequencies and optimizing pulse durations, our method enables the detection of parity switches during measurements.
% What we found:
 By embedding parity detection within Ramsey measurements, we quantitatively link parity switching events to dephasing, establishing charge-parity noise as a dominant decoherence mechanism in this regime. Our combined experimental and numerical results demonstrate that residual charge dispersion in the transmon regime, remains a significant source of $T_2^*$ fluctuations.

% How do we know 
To substantiate this conclusion, we compared spin-echo and Ramsey experiments, as these protocols probe different spectral components of qubit frequency fluctuations. Parity switches occurring on millisecond timescales generate low-frequency noise, which has a measurable impact on $T_2^*$ in Ramsey experiments. The spin-echo sequence, which filters out slow fluctuations, yields stable dephasing times across varying charge dispersion, confirming that the observed $T_2^*$ fluctuations originate from slow parity switching. Furthermore, the relatively slow rate of these switches not only affects coherence but also allows us to directly observe individual parity switching events within our experimental protocol. If parity switches were much faster, we would expect $T_2^{\text{Echo}}$ to decrease with increasing charge dispersion and our ability to resolve individual switches to be compromised.

% Another insights
Additionally, we observed longer-term drifts in $T_2^*$ that cannot be explained by charge-parity noise alone. While our protocol accounts for short-term fluctuations, these slower changes point to additional, unidentified mechanisms affecting qubit coherence over months. Nevertheless, the phenomenon of charge-parity-induced dephasing remains present and consistent across measurements.

% Outlook
Looking forward, this protocol can be extended to other superconducting qubit noise mechanisms to test for any form of charge sensitivity, guided by the intuition developed here regarding the relevant noise regimes. Investigating the interplay between quasiparticle dynamics and other decoherence mechanisms could further inform strategies to optimize coherence in superconducting qubits even if the lowest energy levels of the qubit are charge insensitive.

% A goodbye!
Overall, our results underscore the need to address charge-parity noise as a key factor in the pursuit of longer-lived and more reliable transmon qubits. This could motivate building devices deeper into the transmon regime (higher $E_J/ E_C$), qubits with lower parity switching rates, or applying post-processing techniques like the one presented here to mitigate the effects of charge-parity noise.

\section*{Acknowledgments}
We would like to acknowledge Sayan Patra's assistance with editing and revising this article. This research is supported by the U.S. Department of Energy, Basic Energy Sciences, under award DE-SC0020313. This work was performed under the auspices of the U.S. Department of Energy by Lawrence Livermore National Laboratory under the Contract DE-AC52-07NA7344. Zaman and Wong also thank the support of the National Science Foundation under Grant No. 2125906. LLNL-JRNL-2003358

\appendix
\renewcommand{\thefigure}{\Alph{section}.\arabic{figure}}
\setcounter{figure}{0}
\section{Appendix}\label{sec:App_Lindblad}

\subsection{Lindblad Master equation to simulate charge-parity noise}
To support our measurements, we modeled the system dynamics using the Lindblad master equation evolution in QuTip \cite{johansson2012qutip}\cite{johansson2013qutip} with and without parity switching in the device using the Lindblad master equation \cite{manzano2020lindblad}.

We start by defining the qubit Hamiltonian in the rotating frame
\begin{equation}
    H{\pm} = -\frac{1}{2} \alpha \  a^{\dagger} a^{\dagger} a a  + \delta \omega_{\pm} a^{\dagger} a .
\end{equation}
Here, $\alpha$ is the system's anharmonicity and $\delta \omega_{\pm} = \delta \omega \pm 2\pi\Delta_{01}$ is a parity-dependent detuning. $\delta \omega $ is a control detuning and $\Delta_{01}$ is the charge dispersion in the computational basis consistent with equation (\ref{eq:charge_dispersion}). We introduce the charge-parity noise component through the detuning, since we are modeling the dynamics within the 0-1 manifold.  While higher energy states are affected by parity, we never populate those states in the simulation and so we discount the effects. 

To include the dissipation terms that represent relaxation and dephasing, we define 
\begin{align}
    L_{1} &= \sqrt{\Gamma_1} a \\
    L_{2} &= \sqrt{\Gamma_{\phi}} a^{\dagger} a .   
\end{align}

In this context, 
$$
\Gamma_1 = \frac{1}{T_{1}}, \quad \Gamma_{\phi} = \frac{2}{T_{\phi}}.
$$

$ T_{1} $ and $ T_{\phi} $ are the relaxation and pure dephasing decay rates.

If the noise is Markovian, we expect to have a transverse decay rate:
$$
\frac{1}{T_2^{*}} = \frac{1}{2 T_1} + \frac{1}{T_{\phi}}
$$
which we can extract by performing a Ramsey interferometry experiment.

However, the presence of the parity flips changes the Hamiltonian from $H^{\pm}$ to $H^{\mp}$ at some time $t_f$ during the qubit evolution.

We define a uniform distribution of times $t_f$ within the experiment's duration, where the Hamiltonian transitions from one parity configuration to the other. Depending on the chosen decay experiment, we obtain for each flip time $t_f$ a description of the density matrix $\rho(t)$. 
Averaging over an ensemble of all the realizations of $t_f$, we obtain the desired density matrix that resembles the dynamics of a transmon that suffers from quasi-particle-induced parity flips. Then its relevant components are used to fit $T_{2}^*$.

By comparing these values to those of a static Hamiltonian (no changes in the detuning during the evolution), we observed a reduction in the dephasing time.

We perform these simulations for a range of $\Delta_{01}$ frequency offsets, and extract the fitted $T_2^{*}$ from each Ramsey decay. 
\subsection{Relevance of the charge-parity noise}\label{}

\label{subsec:charge_parity_noise}
We explore the regime in which the impact of charge parity noise becomes relevant. We plot the measured $T_2^{*}$ as a function of $T_2^{*\text{ideal}}$, where the former is obtained from Ramsey simulations that experience parity flips, and the latter remained in a single parity throughout the evolution. We plot this relationship for different charge dispersion ($\Delta_{01}$) values, presented as different curves.

The gray dashed line shows the threshold where the two quantities ($T_2^*$ and $T_2^{* \text{ideal}}$) converge, helping to visualize when charge-parity noise has a measurable effect on qubit dephasing.

These results illustrate that as $\Delta_{01}$ decreases from 6~kHz to 1.5~kHz, the influence of charge-parity noise on qubit dephasing is significantly reduced, causing the measured $T_2^{*}$ to approach the ideal value (i.e., the value obtained in the absence of parity flips). This trend is also evident in both plots of Fig.~\ref{fig:simulation-vs-experiment}, where the Ramsey experiment results increasingly align with the charge-insensitive metric as $\Delta_{12}$, and consequently $\Delta_{01}$, decrease. Overall, these observations confirm that charge-parity noise emerges as a primary limitation for $T_2^{*}$ when charge dispersion is high.

Another factor to consider is that $T_2^{*\text{ideal}}$ can be limited by other decoherence mechanisms, such as energy relaxation ($T_1$ processes). In the regime where $T_2^{*}$ is intrinsically short due to these mechanisms, the impact of charge-parity noise becomes negligible, even if the charge dispersion is high. In other words, when $T_2^{*}$ is limited by factors other than charge-parity noise, the measured and ideal values will be nearly identical regardless of $\Delta_{01}$.

\begin{figure}[ht]
    %\centering
    \includegraphics[width=\linewidth]{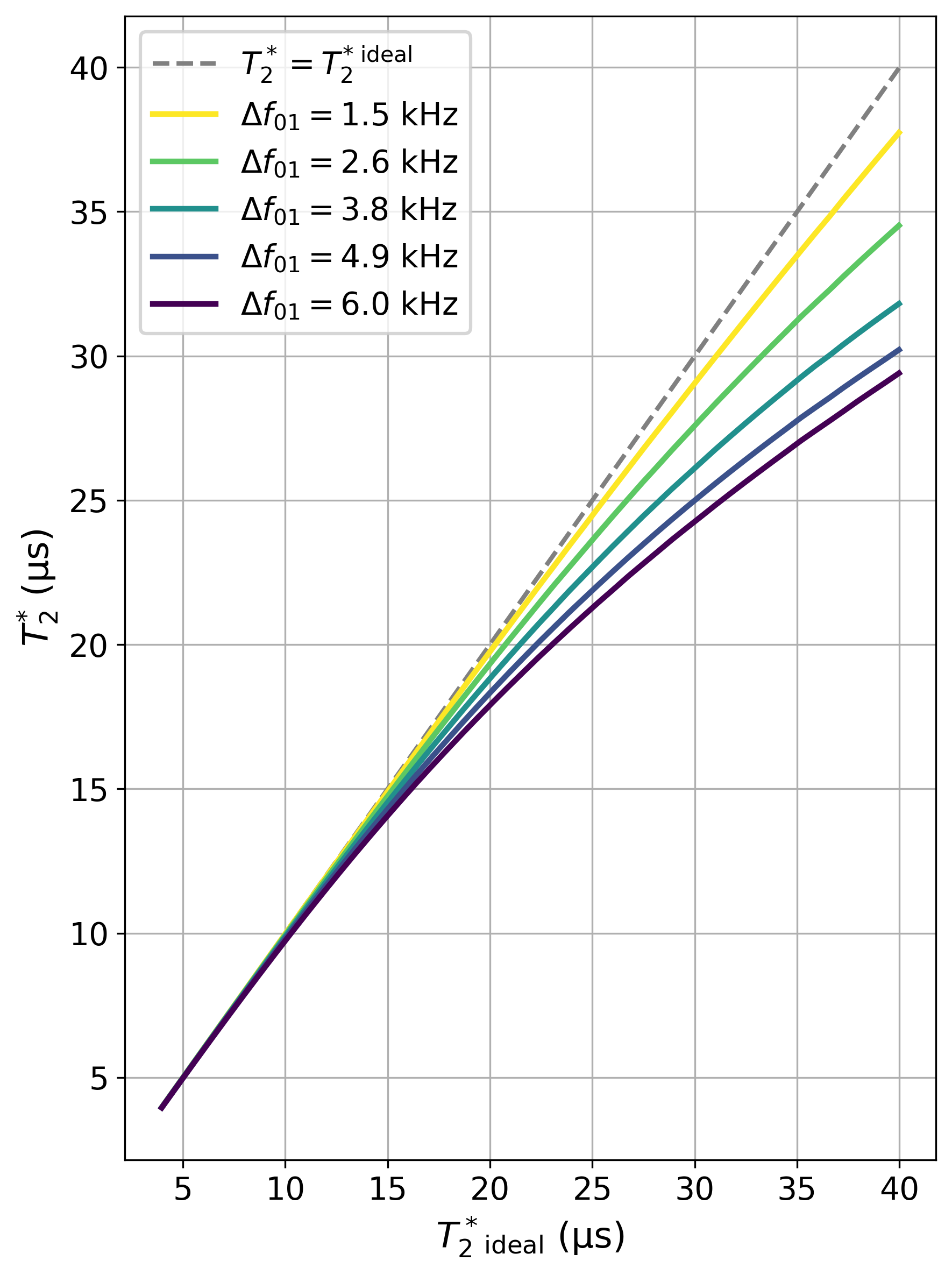}
    \caption{
        Relationship between the measured coherence time $T_2^{*}$ in the presence of charge-parity noise and $T_2^{*\text{ideal}}$ assuming a fixed parity Hamiltonian. Each colored curve corresponds to a different charge offset ($\Delta_{01}$), and the gray dashed line represents the identity ($T_2^* = T_2^{*\text{ideal}}$).
        The deviation between the ideal and measured coherence times becomes more pronounced as $T_2^{*\text{ideal}}$ increases, especially for larger charge dispersion. Since our device operates in the range $T_2^{*\text{ideal}} \sim 38$–$45\,\mu$s, charge-parity noise significantly impacts coherence at these timescales. Conversely, when $T_2^{*\text{ideal}}$ is short, the influence of this noise is negligible.}
    \label{fig:T2star_vs_T2ideal}
\end{figure}

In our experiments, the extracted $T_2^*$ values range from $38~\mu\mathrm{s}$ to $23~\mu\mathrm{s}$ (see Fig.~\ref{fig:simulation-vs-experiment}). According to the analysis presented in Fig.~\ref{fig:T2star_vs_T2ideal}, this range falls well within the regime where charge-parity noise significantly impacts qubit coherence. Consequently, when our simulation allows for the Hamiltonian to rest in one parity frequency, we observe a substantial improvement in $T_2^*$, confirming that charge-parity noise is a dominant decoherence mechanism in our devices.

\bibliographystyle{unsrt}
\bibliography{citations}

\end{document}